# Ultracompact and Nonvolatile On-chip Switch Based on DBS and PSO Algorithms

## Sanaz Zarei

Sharif University of Technology, Tehran, Iran

szarei@sharif.edu

***Abstract*—A 1×2 photonic switch based on silicon-on-insulator (SOI)-$Sb_2Se_3$ hybrid platform is inverse-designed using direct binary search (DBS) and particle swarm optimization (PSO) algorithms. The nonvolatile on-chip switch is designed to consist of minimum switchable phase-change units and operates based on the en-masse switching of all $Sb_2Se_3$ cells from crystalline to amorphous and vice-versa. The device has a small footprint of only 6.3μm×31μm. At the wavelength of 1550nm, it exhibits an insertion loss of less than 0.96dB, and its crosstalk is measured to be less than -12dB within the three-dimensional (3D) finite-difference time-domain (FDTD) simulations.**

## 1. Introduction

On-chip photonic integrated circuits (PICs) based on silicon-on-insulator (SOI) platform, with the advantages like broad bandwidth, low power consumption, and low loss, have enabled proficient optical signal control and processing such as modulation, switching, and filtering [1]. Among them, silicon photonic switches are pivotal building blocks that construct programmable PICs and facilitate optical signal routing and switching [2]. Therefore, a photonic switch with a small footprint, low loss, low energy consumption, low crosstalk, and large bandwidth, will dramatically assist in the development of programmable PICs [2]. The state-of-the-art photonic switches are mainly based on the Mach-Zehnder interferometers (MZI) or micro-ring resonators (MRR). The refractive index adjustment in such structures using free-carrier dispersion or thermo-optic effects suffers from a small change in the complex refractive index, which necessitates larger dimensions for the devices [2]. Furthermore, they are volatile, and continuous input energy is required for maintaining their switching states, which increases their power consumption [2].

Phase-change materials (PCMs) create much more substantial refractive index contrast than electro-optic or thermo-optic effects and are widely used in optical switches [1-5]. The sulfur compound $Sb_2Se_3$ is a wide bandgap PCM with a transparency window from 600nm to 1700nm [1]. Its refractive index is 3.285 + 0i in the amorphous state and 4.05 + 0i in the crystalline state [1]. These characteristics make $Sb_2Se_3$ highly suitable for application in programmable PICs.

Recently, inverse design algorithms have been successful in designing ultracompact and highly functional devices simultaneously. Several algorithms have been presented so far, including direct binary search (DBS) algorithms [1-3, 6], particle swarm optimization (PSO) algorithms [7], genetic algorithms [8-9], and deep learning [10]. Here, a compact and nonvolatile on-chip switch based on $Sb_2Se_3$ metasurfaces is inverse-designed using DBS and PSO algorithms and the finite-difference time-domain (FDTD) method.

## 2. On-chip Switch Architecture

In this work, a standard 220nm-thick SOI platform is considered for the realization of the on-chip diffractive switch. The switch body, waveguides and tapers can be created by total etching of the silicon (Si) device layer. The architecture of the on-chip switch is shown in Fig. 1(a). The 1×2 on-chip switch with a footprint of 6.3μm×31μm, consists of five phase-change metasurfaces [8, 11-14], each having 21 meta-atoms. The first on-chip one-dimensional (1D) metasurface is located at x=0 and all other 1D metasurfaces are located sequentially with 6μm separations throughout the metasystem. The input and outputs of the metasystem are fixed at x=-1μm and x=30μm, respectively. The input/output waveguides are 0.9μm wide and the input/output tapers have 1.8μm width when in conjunction with the switch body and 5μm length. The center-to-center distance between the output waveguides/tapers is 3μm. A phase-change metasurface is a one-dimensional array of $Sb_2Se_3$ rods that are created inside the silicon-on-insulator (SOI) substrate. A 2μm-thick $SiO_2$ upper cladding exists on the silicon layer for protection and diffusion inhibition of $Sb_2Se_3$ and the underneath $SiO_2$ insulator layer also has 2μm thickness. The phase change metasurfaces form different wavefronts when the $Sb_2Se_3$ inclusions, en-masse, switch between the amorphous and crystalline states. The $Sb_2Se_3$ rod arrays are parametrized by the lattice constant of the 1D arrays (a) and length (L), width (W), and height (H) of rods. While the lattice constant of the arrays and height of the rods are fixed at 300nm and 220nm, respectively, the in-plane geometric parameters of length and width of each subwavelength rod are chosen as the design parameters and are swept to achieve the transmission amplitude and phase map of each rod in both amorphous (Fig. 1(b) and Fig. 1(c)) and crystalline states (Fig. 1(d) and Fig. 1(e)), for the fundamental TE-polarized guided wave. For the sake of easier phase change and lower loss, the width of the rods is constantly chosen as W=100nm. By this selection, in Fig. 1(f), the electric field transmission phase and amplitude of a $Sb_2Se_3$ rod is calculated for the fundamental TE-polarized guided wave, by changing the rod length from 0 to 5µm at both amorphous and crystalline states. For both states, the transmission amplitude is very near to 1 which is due to the ultralow-loss of $Sb_2Se_3$ and the proximity of its refractive index to that of silicon. As can be observed, a rod length of 4µm provides more than π phase difference between amorphous and crystalline states. The electromagnetic simulation results in Fig. 1 are calculated using finite difference time domain algorithm (Lumerical FDTD). For these calculations, FDTD ports with a fixed distance of 6µm along the x-direction, x-axis as the injection axis, and 1.55µm as the excitation wavelength, are utilized.

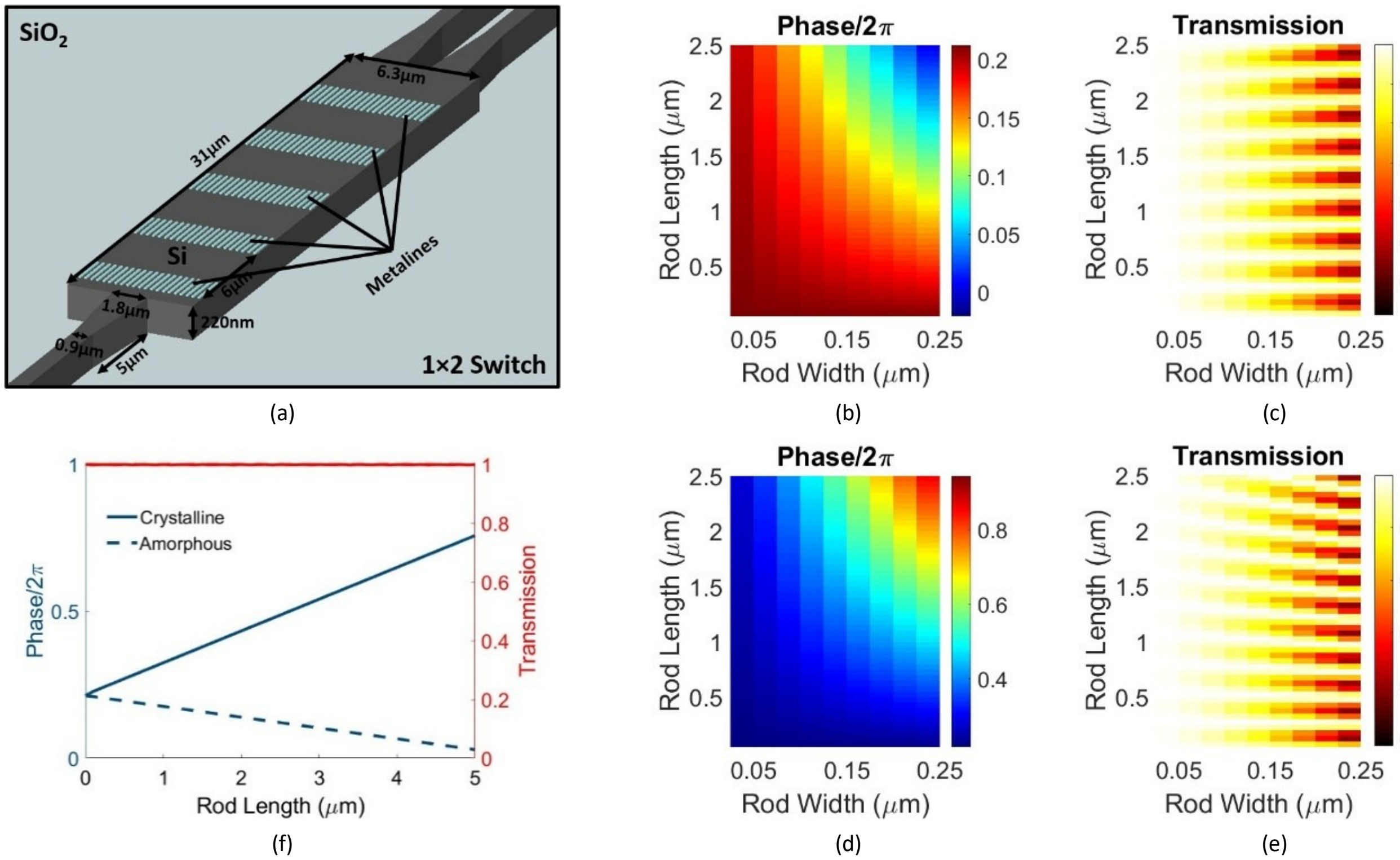

Fig. 1. (a) Schematic of a 1×2 on-chip switch within a SOI platform consisting of five 1D phase-change metasurfaces. Electric field transmission of TE-polarized guided wave for a $Sb_2Se_3$ rod versus rod length and rod width for: (b) and (c) $Sb_2Se_3$ in amorphous state, (d) and (e) $Sb_2Se_3$ in crystalline state. (b) and (d) illustrate the transmission phase, and (c) and (e) represent the transmission amplitude. (f) Electric field transmission phase and amplitude versus $Sb_2Se_3$ rod length, fixing the rod width at 100nm. The operation wavelength is set to 1.55µm. These diagrams are produced by the commercial software package Lumerical FDTD.

## 3. On-chip Switch Design

The design of the on-chip diffractive switch is accomplished in three rounds. In the first round, an optimization based on the direct binary search (DBS) algorithm is applied to the initial configuration shown in Fig. 1(a), which is composed of five phase-change metasurfaces in which the length of all $Sb_2Se_3$ rods is set to 2µm. Throughout the DBS algorithm, each rod within the phase-change metasurfaces, can be filled with either Si or $Sb_2Se_3$. This process, which can be interpreted as sparse parameterization, aims to drop unnecessary $Sb_2Se_3$ rods and decrease the number of phase-change cells, while maximizing the performance. A figure of merit (FOM) is defined to evaluate the optimized performance. The rods are serially switched from $Sb_2Se_3$ to Si or vice versa and for each change, the FOM is calculated for both amorphous and crystalline states. If the FOM decreases, the material change will be retained. Otherwise, it will return to its original material. The FOM is defined as follows:

$$\begin{aligned} FOM &= FOM_{amorphous} + FOM_{crystalline} \\ FOM_{amorphous} &= \left|1-T_1\right| + \left|T_2\right| \\ FOM_{crystalline} &= \left|T_1\right| + \left|1-T_2\right| \end{aligned} \tag{1}$$

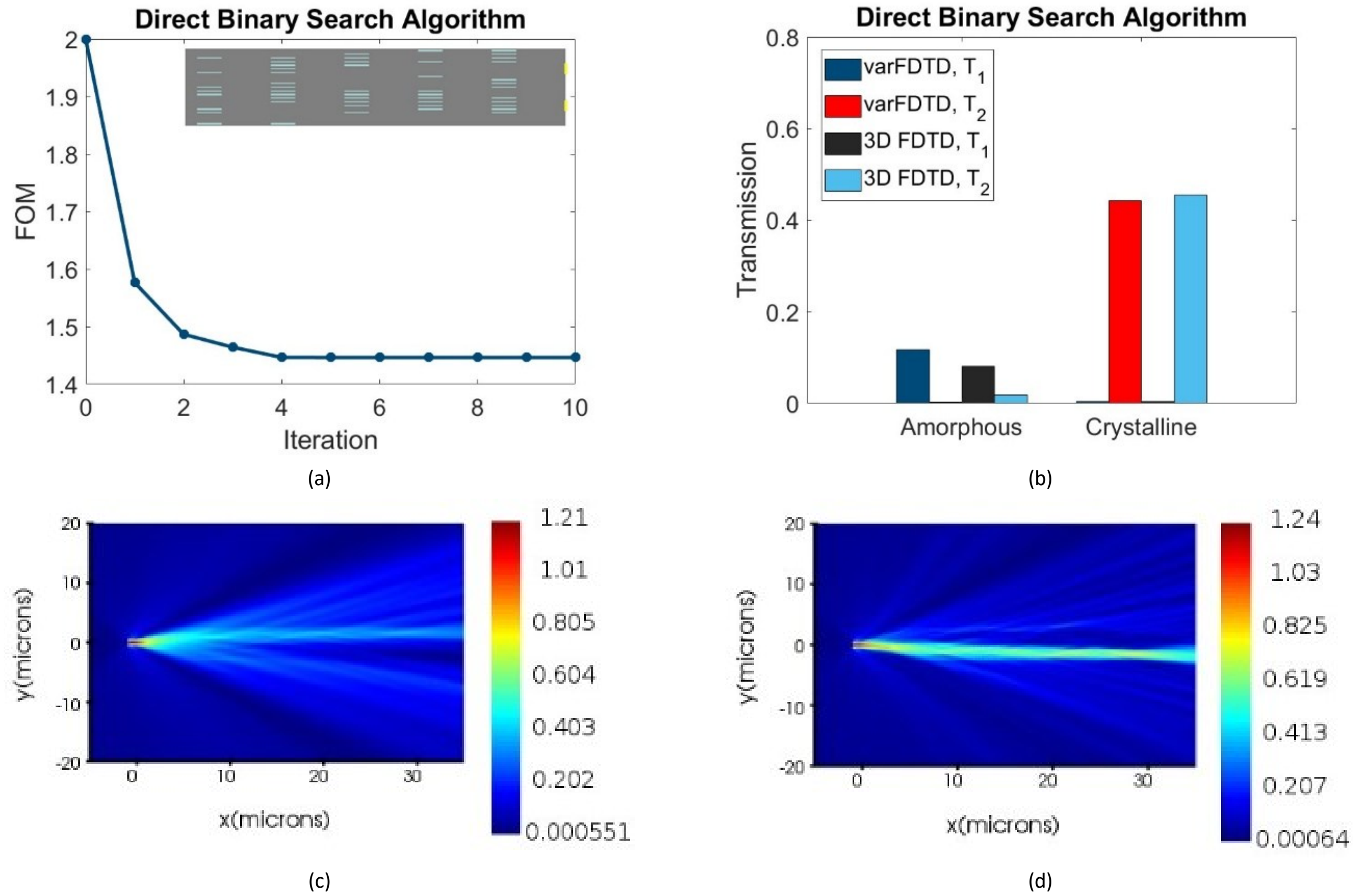

Fig. 2. (a) The FOM change versus iteration throughout the DBS optimization; the inset presents the optimized metasystem. (b) The output (absolute) transmissions at both amorphous and crystalline states in the optimized metasystem, calculated by Lumerical Mode Solution 2.5D variational FDTD solver (varFDTD) and verified by 3D FDTD. The x-y view of the electric field distribution in the optimized metasystem, at the midplane of the silicon device layer (c) in amorphous state, (d) in crystalline state.

where $T_1$ is the transmittance from switch body to the upper taper/waveguide and $T_2$ is the transmittance from the switch body to the lower taper/waveguide (see Fig. 4(a)). The defined FOM targets at maximizing $T_1$ and minimizing $T_2$ in the amorphous state, while minimizing $T_1$ and maximizing $T_2$ in the crystalline state. As is seen in Fig. 2(a), the initial FOM (that belongs to the configuration in Fig. 1(a)) is equal to 2 and as the optimization proceeds, it reduces to 1.44627 at the fifth iteration. No further improvement happens afterwards and the optimization is terminated after the tenth iteration. The inset of Fig. 2(a) illustrates the optimized metasystem and Fig. 2(b) presents the optimized transmissions of the outputs at both states. Figures 2(c) and 2(d) depict the x-y view of the electric field distribution in the optimized metasystem, at the midplane of the silicon device layer in amorphous and crystalline states, respectively.

In the second round of design, particle swarm optimization (PSO) is applied to the optimized structure from the previous round and the FOM is further minimized to 1.1721 in 100 iterations (Fig. 3(a)). Throughout the optimization, the lengths of all existing $Sb_2Se_3$ rods are adjusted within the range of 300nm and 4.5µm to minimize the FOM. Starting from the initial configuration, each particle travels the parameter space to find the best configuration. The FOM is again set as the equation 1. The optimum metasystem is pictured in the inset of Fig. 3(a), Fig. 3(b) illustrates the outputs' transmissions in this configuration at both amorphous and crystalline states, and figures 3(c) and 3(d) pictures the x-y view of the electric field distribution in the metasystem, at the middle plane of the silicon layer in amorphous and crystalline states, respectively.

For both optimizations, the 2.5D variational FDTD solver (varFDTD) of Lumerical Mode Solution is utilized for structural modeling and the obtained results are verified with Lumerical 3D FDTD (Fig. 2(b) and Fig. 3(b)). It should be noted that both design rounds were performed within a metasystem composed of five phase-change metasurfaces in the SOI platform without the presence of tapers and waveguides and any confinement imposed by the switch body. This is because the varFDTD simulation time for such metasystem is much shorter than the simulation time of the complete 1×2 switch module shown in Fig. 1(a). Throughout the simulations, instead of the input waveguide and taper, a mode source of 0.9µm width, located at x=-1µm, is used for excitation and instead of output tapers and waveguides, two output monitors of 0.9µm width and 3µm center-to-center distance were exploited for electric field transmission measurements.

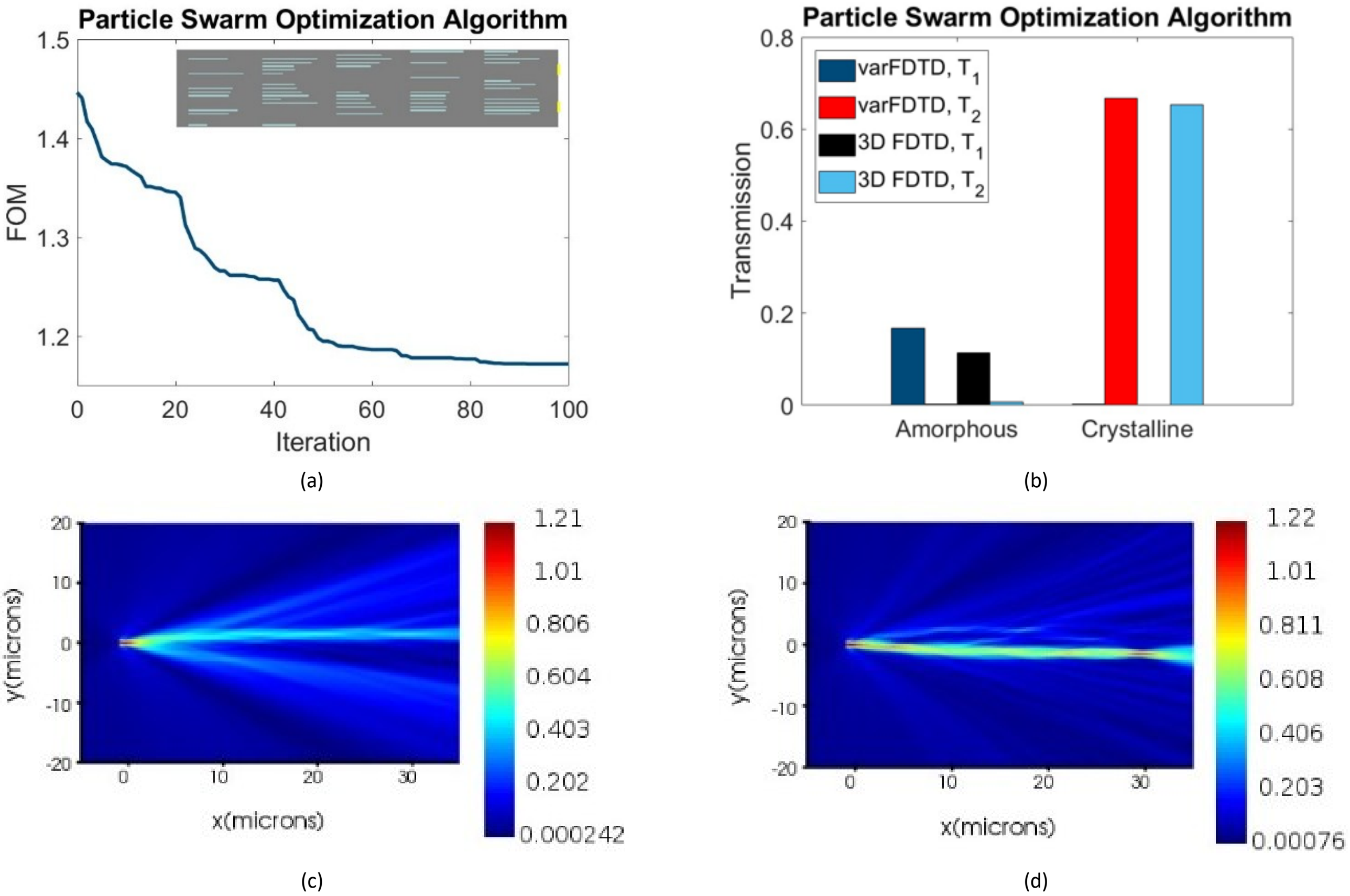

Fig. 3. (a) The FOM change versus iteration throughout the PSO optimization; the inset presents the optimized metasystem. (b) The output (absolute) transmissions of the optimized metasystem at both amorphous and crystalline sates, calculated by varFDTD and verified by 3D FDTD. The x-y view of the electric field distribution in the optimized metasystem, at the midplane of the silicon device layer (c) in amorphous state, (d) in crystalline state.

By transferring the finally optimized metasystem into the 1×2 switch module shown in Fig. 1(a), the total transmission of the system enhances (the FOM decreases to 0.7395), however, the performance of the system as a switch degrades, because of the increment of transmission through each output taper/waveguide at both states. Therefore, the third round of optimization is applied to fine tune the metasystem and restore the performance. Again, PSO is employed for this design round and the optimization with 80 iterations (Fig. 4(a)) lasts much longer due to the time-consuming nature of such metasystem simulations. The ultimate optimized metasystem is presented in the inset of Fig. 4(a). Figure 4(b) shows the optimized transmissions at both states (calculated by both varFDTD and 3D FDTD) and figures 4(c) and 4(d) illustrate the x-y view of the electric field distribution in the optimized metasystem at both states.

As is illustrated in Fig. 4(b), the insertion loss (IL) decreases comparing to the optimized metasystems in figures 2(b) and 3(b), but the crosstalk (CT) increases, especially at amorphous state. The final ILs calculated from the following equation at both amorphous and crystalline states are 0.96dB and 0.45dB (based on 3D FDTD results), respectively. Also, the computed CTs based on equation (3) are approximately -12dB at amorphous state and -15dB at crystalline state (calculated from 3D FDTD results).

$$IL = -10\log(T_{1(2)} / T_{in}) \tag{2}$$

$$CT = 10\log(T_{2(1)} / T_{1(2)}) \tag{3}$$

where $T_{in}$ is the transmittance from the input taper/waveguide to the switch body. It is expected that better performance can be obtained by performing all optimizations within the 1×2 switch module from the beginning. However, such optimizations will take a long time. As a final remark, the idea behind the utilization of DBS for initial optimization was to decrease the number of phase-change cells to ease fabrication and facilitate switching, however, better IL and CT may be attained by performing PSO optimization on all the phase-change cells, which increases the number of degrees of freedom through the design process. In the presented design, the number of phase-change cells are 51 out of 105 for all the optimized configurations.

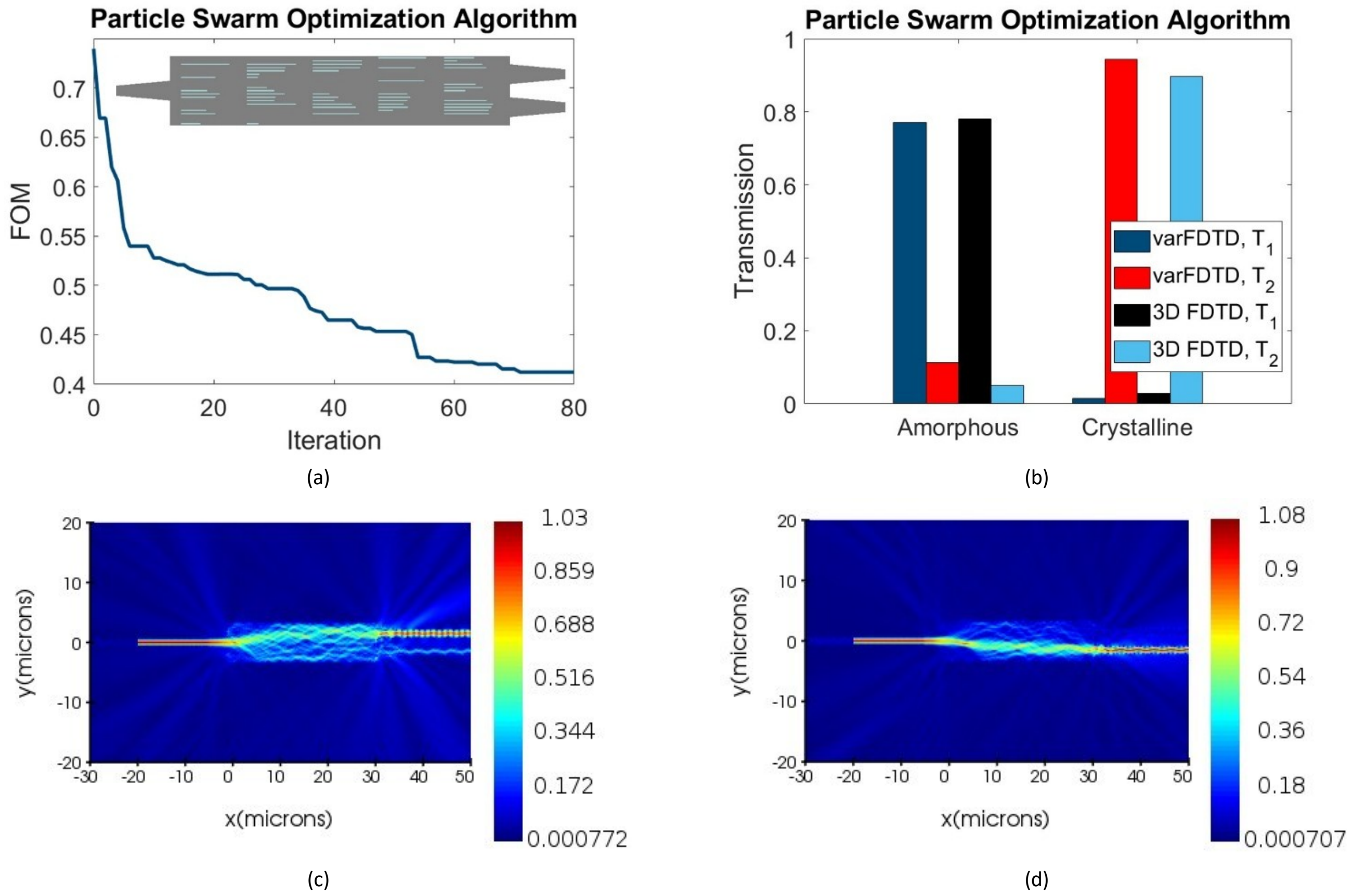

Fig. 4. (a) The FOM change versus iteration throughout the final PSO optimization and the Schematic of the finally-optimized 1×2 on-chip diffractive switch in the inset, (b) The output (absolute) transmissions of the optimized switch at both amorphous and crystalline states, calculated by both varFDTD and 3D FDTD. The x-y view of the electric field distribution in the optimized switch, at the midplane of the silicon device layer (c) in amorphous state, (d) in crystalline state.

## 4. Conclusions

In conclusion, an on-chip 1×2 diffraction-based switch, founded on the en-masse switching of the $Sb_2Se_3$ cells within the switch body between the amorphous and crystalline states, is presented. The presented nonvolatile switch has a very small footprint of 6.3µm×31µm. A combination of DBS and PSO algorithms was exploited to optimize the switching performance of the device and FDTD was applied for device modeling. Based on 3D FDTD simulations, low insertion loss and crosstalk of less than 0.96dB and -12dB, respectively, has been demonstrated for the device. Further optimization of the switch performance is a future work to be accomplished.

## Statements & Declarations

### Funding

The author declares that no funds, grants, or other support were received during the preparation of this manuscript.

### Competing Interests

The author has no relevant financial or non-financial interests to disclose.

### Author Contributions

S. Z. performed the study conception and design, accomplished the FDTD simulations, carried out the DBS and PSO optimizations, analyzed the results, and wrote the manuscript.

### Data Availability

No datasets were generated or analyzed during this study.

### Ethics declarations

This article does not contain any studies involving animals or human participants.